# Predicting article quality scores with machine learning: The UK Research Excellence Framework

Mike Thelwall
Statistical Cybermetrics and Research Evaluation Group, University of Wolverhampton, UK.
https://orcid.org/0000-0001-6065-205X m.thelwall@wlv.ac.uk

Kayvan Kousha
Statistical Cybermetrics and Research Evaluation Group, University of Wolverhampton, UK.
https://orcid.org/0000-0003-4827-971X k.kousha@wlv.ac.uk

Paul Wilson
Statistical Cybermetrics and Research Evaluation Group, University of Wolverhampton, UK.
https://orcid.org/0000-0002-1265-543X pauljwilson@wlv.ac.uk

Meiko Makita
Statistical Cybermetrics and Research Evaluation Group, University of Wolverhampton, UK.
https://orcid.org/0000-0002-2284-0161 meikomakita@wlv.ac.uk

Mahshid Abdoli
Statistical Cybermetrics and Research Evaluation Group, University of Wolverhampton, UK.
https://orcid.org/0000-0001-9251-5391 m.abdoli@wlv.ac.uk

Emma Stuart
Statistical Cybermetrics and Research Evaluation Group, University of Wolverhampton, UK.
https://orcid.org/0000-0003-4807-7659 emma.stuart@wlv.ac.uk

Jonathan Levitt
Statistical Cybermetrics and Research Evaluation Group, University of Wolverhampton, UK.
https://orcid.org/0000-0002-4386-3813 j.m.levitt@wlv.ac.uk

Petr Knoth
Knowledge Media Institute, Open University, UK
https://orcid.org/0000-0003-1161-7359 petr.knoth@wlv.ac.uk

Matteo Cancellieri
Knowledge Media Institute, Open University, UK
https://orcid.org/0000-0002-9558-9772 matteo.cancellieri@open.ac.uk

National research evaluation initiatives and incentive schemes have previously chosen between simplistic quantitative indicators and time-consuming peer review, sometimes supported by bibliometrics. Here we assess whether artificial intelligence (AI) could provide a third alternative, estimating article quality using more multiple bibliometric and metadata inputs. We investigated this using provisional three-level REF2021 peer review scores for 84,966 articles submitted to the UK Research Excellence Framework 2021, matching a Scopus record 2014-18 and with a substantial abstract. We found that accuracy is highest in the medical and physical sciences Units of Assessment (UoAs) and economics, reaching 42% above the baseline (72% overall) in the best case. This is based on 1000 bibliometric inputs and half of the articles used for training in each UoA. Prediction accuracies above the baseline for the social science, mathematics, engineering, arts, and humanities UoAs were much lower or close to zero. The Random Forest Classifier (standard or ordinal) and Extreme Gradient Boosting Classifier algorithms performed best from the 32 tested. Accuracy was lower if UoAs were merged or replaced by Scopus broad categories. We increased accuracy with an active learning strategy and by selecting articles with higher prediction probabilities, as estimated by the algorithms, but this substantially reduced the number of scores predicted.

# 1 Introduction

Many countries systematically assess the outputs of their academic researchers to monitor progress or reward achievements. A simple mechanism for this is to set bibliometric criteria to gain rewards, such as awarding funding for articles with a given Journal Impact Factor (JIF). Several nations have periodic evaluations of research units instead, however. These might be simultaneous nationwide evaluations (e.g., Australia, New Zealand, UK: Buckle & Creedy, 2019; Hinze, et al. 2019; Wilsdon et al., 2015a), or rolling evaluations for departments, fields, or funding initiatives (e.g., The Netherlands' Standard Evaluation Protocol: Prins et al., 2016). Peer review seems to be the most desirable system because reliance on bibliometric indicators can disadvantage some research groups, such as those that focus on applications rather than theory or methods development. Nevertheless, extensive peer review requires a substantial time investment from experts with the skill to assess academic research quality and there is a risk of human bias, which are major disadvantages. In response, some systems inform peer review with bibliometric indicators (UK: Wilsdon et al., 2015) or automatically score outputs that meet certain criteria, reserving human reviewing for the remainder (as Italy used to: Franceschini & Maisano, 2017). In this article we assess a third approach: machine learning to estimate the score of some outputs in a periodic research assessment, as previously proposed (Thelwall, 2022). It is evaluated for the first time here with peer review scores for a large set of articles.

The background literature relevant to predicting article scores with machine learning has been introduced in a prior article (Thelwall, 2022) that also reported experiments with Artificial Intelligence (AI) applied to predict the citation rate of an article's publishing journal as a proxy article quality measurement. This study found that the Gradience Boosting Classifier was the most accurate out of 32 classifiers tested. Its accuracy varied substantially between the 326 Scopus narrow fields that it was applied to, but it had above chance accuracy in all fields. The text inputs into the algorithm seemed to leverage journal-related style and boilerplate text information rather than more direct indicators of article quality, however. The level of error in the results was large enough to generate substantial differences between institutions through changed average scores. Another previous study had used simple statistical regression to predict REF scores for individual articles in the 2014 REF, finding that the value of journal impact and article citation counts varied substantially between Units of Assessment (UoAs) (HEFCE, 2015).

Although comparisons between human scores and computer predictions for journal article quality tend to assume that the human scores are correct, even experts are likely to disagree and can be biased. An underlying reason for disagreements is that many aspects of peer review are not well understood, including the criteria that articles should be assessed against (Tennant & Ross-Hellauer, 2020). For REF2014 and REF2021, articles were assessed for originality, significance, and rigour (REF2021, 2019), which is the same as the Italian Valutazione della Qualità della Ricerca requirement for originality, impact, and rigour (Bonaccorsi, 2020). These criteria are probably the same as for most journal article peer reviewing, except that some journals mainly require rigorous methods (e.g., PLOS, 2022). Bias in peer review can be thought of as any judgment that deviates from the true quality of the article assessed, although this is impossible to measure directly (Lee et al., 2013). Non-systematic judgement differences are also common (Jackson et al., 2011; Kravitz, et al., 2010). Non-systematic differences may be due to unskilled reviewers, differing levels of leniency or experience (Haffar et al., 2019; Jukola, 2017), weak disciplinary norms (Hemlin, 2009), and perhaps also due to teams of reviewers focusing on different aspects of a paper (e.g., methods, contribution, originality). Weak disciplinary norms can occur because a field's research objects/subjects and methods are fragmented or because there are different schools of thought about which theories, methods or paradigms are most suitable (Whitley, 2000).

Sources of systematic bias that have been suggested for non-blinded peer review include malicious bias or favouritism towards individuals (Medoff, 2003), gender (Morgan, et al., 2018), nationality (Thelwall et al., 2021), and individual or institutional prestige (Bol, et al.., 2018). Systematic peer review bias may also be based on language (Herrera, 1999; Ross et al., 2006), and study topic or approach (Lee et al., 2013). There can also be systematic bias against challenging findings (Wessely, 1998), complex methods (Kitayama, 2017), or negative results (Gershoni et al., 2018). Studies that find review outcomes differing between groups may be unable to demonstrate bias rather than other factors (e.g., Fox & Paine, 2019). For example, worse peer review outcomes for some groups might be due to lower quality publications because of limited access to resources, or unpopular topic choices. A study finding some evidence of same country reviewer systematic bias that accounted for this difference could not rule out the possibility that it was a second order effect due to differing country specialisms and same-specialism systematic reviewer bias rather than national bias (Thelwall et al., 2021).

In this article we assess whether it is reasonable to use machine learning to estimate any UK REF output scores for journal articles. It is a condensed and partly rephrased version of a longer report, with some additional analyses. We detail three approaches: (a) human scoring for a fraction of the outputs, then AI predictions for the remainder; (b) human scoring for a fraction of the outputs, then AI predictions for a subset of the rest where the predictions have a high probability of being correct; and human scoring for the remaining articles; and (c) the active learning strategy to identify sets of articles that meet a given probability threshold. These are assessed with expert peer review quality scores for most of the journal articles submitted to REF2021. The research questions are as follows, with the final research question introduced to test if the results change with a different standard classification schema.
- RQ1: How accurately can AI estimate article quality from article metadata and bibliometric information in each scientific field?
- RQ2: Which machine learning methods are the most accurate for predicting article quality in each scientific field?
- RQ3: Can higher accuracy be achieved on subsets of articles using AI prediction probabilities or active learning?
- RQ4: How accurate are AI article quality estimates when aggregated over institutions?
- RQ5: Is the AI accuracy similar for articles organised into Scopus broad fields?

## 2 Methods

The research design was to assess a range of AI algorithms in a traditional training/testing validation format: training each algorithm on a subset of the data and evaluating it on the remaining data.

### 2.1 Data: Articles and scores

We used data from two data sources. First, we downloaded records for all Scopus-indexed journal articles published 2014-2020 in January-February 2021 using the Scopus Applications Programming Interface (API). This matches the date when the human REF2021 assessments were originally scheduled to begin, so is from the time frame when a machine learning stage could be activated. We excluded reviews and other non-article records in Scopus for consistency. The second source was a set of 148,977 provisional article quality scores assigned by the expert REF sub-panel members to the articles in 34 UoAs, excluding all data from the University of Wolverhampton. This was confidential data that could not be shared and had to be deleted before 10 May 2022. The REF data included article DOIs (used for matching with Scopus, and validated by the REF team), evaluating UoA (one of 34), and provisional score (0, 1*,2*,3*, or 4*). We merged the REF scores into three groups for analysis: 1 (0, 1* and 2*); 2 (3*) and 3 (4*). The grouping was necessary because there were few articles with scores of 0 or 1*, which is problematic for machine learning. This is a reasonable adjustment because 0, 1* and 2* all have the same level of REF funding (zero), so they are financially equivalent.

We matched the REF outputs with journal articles in Scopus with a registered publication date from 2014 to 2020 (Table 1). Matching was primarily achieved through DOIs. We checked papers without a matching DOI in Scopus against Scopus by title, after removing non-alphabetic characters (including spaces) and converting to lowercase. We manually checked title matches for publication year, journal name, and author affiliations. When there was a disagreement between the REF registered publication year and the Scopus publication year, we always used the Scopus publication year. The few articles scoring 0 appeared to be mainly anomalies, seeming to have been judged unsuitable for review due to lack of evidence of substantial author contributions or being an inappropriate type of output. We therefore excluded these.

Finally, we also examined a sample of articles without an abstract. For the five fields with the highest AI accuracy in preliminary tests, these were mainly short format (e.g., letters, communications) or nonstandard articles (e.g., guidelines), although data processing errors accounted for a minority of articles with short or missing abstracts. Short format and unusual articles seem likely to be difficult to predict with AI and so we excluded articles with abstracts shorter than 500 characters. Thus, the final set was cleansed of articles that could be identified in advance as unsuitable for AI predictions. The most accurate predictions were found for the years 2014-18, with at least two full years of citation data, so we reported these for the main analysis as the highest accuracy subset.

Table 1. Descriptive statistics for creation of the experimental dataset.

| Set of articles | Journal articles |
| --- | --- |
| REF2021 journal articles supplied. | 148,977 |
| With DOI. | 147,164 (98.8%) |
| With DOI and matching Scopus 2014-20 by DOI. | 133,218 (89.4%) |
| Not matching Scopus by DOI but matching with Scopus 2014-20 by title. | 997 (0.7%) |
| Not matched in Scopus and excluded from analysis. | 14,762 (9.9%) |
| All REF2021 journal articles matched in Scopus 2014-20. | 134,215 (90.1%) |
| All REF2021 journal articles matched in Scopus 2014-20 except score 0. | 134,031 (90.0%) |
| All non-duplicate REF2021 journal articles matched in Scopus 2014-20 except score 0. | 122,331 [90.0% effective] |
| All non-duplicate REF2021 journal articles matched in Scopus **2014-18** except score 0. These are the most accurate prediction years. | 87,739 |
| All non-duplicate REF2021 journal articles matched in Scopus **2014-18** except score 0 and except articles with less than 500 character cleaned abstracts. | 84,966 |

The 2014-18 articles were mainly from Main Panel A (33,256), Main Panel B (30,354) and Main Panel C (26,013), with a much smaller number from Main Panel D (4,209). The number per sub-panel 2014-18 varied by several orders of magnitude, from 56 (Classics) to 12,511 (Engineering), as shown below in a results table. The number of articles affects the accuracy of machine learning and there were too few in Classics to build machine learning models.

## 2.2 Machine learning inputs

We used textual and bibliometric data as inputs for all the machine learning algorithms. We used all inputs shown in previous research to be useful for predicting citations counts, as far as possible, as well as some new inputs that seemed likely to be useful. We also adapted input used in previous research to use bibliometric best practice, as described below. The starting point was the set of inputs used in a previous citation-based study (Thelwall, 2022) but this was extended. Most of the inputs were tested statistically before the AI model was built to help select the final set.

The citation data for several inputs was based on the Normalised Log-transformed Citation Score (NLCS) or the Mean Normalised Log-transformed Citation Score (MNLCS) (Thelwall, 2017). The NLCS of an article uses log-transformed citation counts, as follows. First, we transformed all citation counts with the natural log ln(1+x). This transformation was necessary because citation count data is highly skewed and taking the arithmetic mean of a skewed dataset can give a poor central tendency estimate. After this, we normalised the log-transformed citation count for each article by dividing by the average log-transformed citation count for its Scopus narrow field and year. We divided articles in multiple Scopus narrow fields instead by the average of the field averages for all these narrow fields. The result of this calculation is an NLCS for each article in the Scopus dataset (including those not in the REF). The NLCS of an article is field and year normalised by design, so a score of 1 for any article in any field and year always means that the article has had average log-transformed citation impact for its field and year. We calculated the following from the NLCS values and used them as machine learning inputs.

- Author MNLCS: The average NLCS for all articles 2014-20 in the Scopus dataset including the author (identified by Scopus ID).
- Journal MNLCS for a given year: The average NLCS for all articles in the Scopus dataset in the specified year from the journal. Averaging log-transformed citation counts instead of raw citation counts gives a better estimate of central tendency for a journal (e.g., Thelwall & Fairclough, 2015).

**Input set 1**: **bibliometrics**. The following nine indicators have been shown in previous studies to associate with citation counts, including readability (e.g., Didegah & Thelwall, 2013), author affiliations (e.g., Fu & Aliferis, 2010; Li et al., 2019a; Qian et al., 2017; Zhu & Ban, 2018), and author career factors (e.g., Qian et al., 2017; Wen et al., 2020; Xu et al., 2019; Zhu & Ban, 2018). We selected the first author for some indicators because they are usually the most important (de Moya-Anegon et al., 2018; Mattsson et al., 2011), although corresponding and last authors are sometimes more important in some fields. We used some indicators based on the maximum author in a team to catch important authors that might appear elsewhere in a list.

1. **Citation counts** (field and year normalised to allow parity between fields and years, log transformed to reduce skewing to support linear-based algorithms).
2. **Number of authors** (log transformed to reduce skewing). Articles with more authors tend to be more cited, so they are likely to also be more highly rated (Thelwall & Sud, 2016).
3. **Number of institutions** (log transformed to reduce skewing). Articles with more institutional affiliations tend to be more cited, so they are likely to also be more highly rated (Didegah & Thelwall, 2013).
4. **Number of countries** (log transformed to reduce skewing). Articles with more country affiliations tend to be more cited, so they are likely to also be more highly rated (Wagner et al., 2019).
5. **Number of Scopus-indexed journal articles of the first author** during the REF period (log transformed to reduce skewing). More productive authors tend to be more cited (Abramo et al., 2014; Larivière & Costas, 2016), so this is a promising input.
6. **Average citation rate of Scopus-indexed journal articles by the first author** during the REF period (field and year normalised, log transformed: the MNLCS). Authors with a track record of highly cited articles seem likely to write higher quality articles. Note that the first author may not be the REF submitting author or from their institution because the goal is not to "reward" citations for the REF author but to predict the score of their article.
7. **Average citation rate of Scopus-indexed journal articles by any author** during the REF period (maximum) (field and year normalised, log transformed: the MNLCS). Again, authors with a track record of highly cited articles seem likely to write higher quality articles. The maximum bibliometric scores in a team has been previously used in another context (Van den Besselaar & Leydesdorff, 2009).
8. **Number of pages of article, as reported by Scopus, or the UoA/Main Panel median if missing from Scopus.** Longer papers may have more content but short papers may be required by more prestigious journals.
9. **Abstract readability**. Abstract readability was calculated using the Flesch-Kincaid grade level score and has shown to have a weak association with citation rates (Didegah & Thelwall, 2013).

**Input set 2**: **bibliometrics + journal impact.** Journal impact indicators are expected to be powerful in some fields, especially for newer articles (e.g., Levitt & Thelwall, 2011). The second input set adds a measure of journal impact to the first set. We used the journal MNLCS instead of Journal Impact Factors (JIFs) as an indicator of average journal impact because field normalised values align better with human journal rankings (Haddawy et al., 2016), probably due to comparability between disciplines. This is important because the 34 UoAs are relatively broad, all covering multiple Scopus narrow fields.

10. **Journal citation rate** (field normalised, log transformed [MNLCS], based on the current year for older years, based on 3 years for 1-2 years' old articles).

**Input set 3**: **bibliometrics + journal impact + text.** The final input set also includes text. Text mining may find words and phrases associated with good research (e.g., a simple formula has been identified for one psychology journal: Kitayama, 2017). Text mining for score prediction is likely to leverage hot topics in constituent fields (e.g., Hu et al., 2020; Thelwall & Sud, 2021), as well as common methods (e.g., Fairclough & Thelwall, 2022; Thelwall & Nevill, 2021; Thelwall & Wilson, 2016), since these have been shown to associate with above average citation rates. Hot topics in some fields tend to be highly cited and probably have higher quality articles, as judged by peers. This would be more evident in the more stable arts and humanities related UoAs but these are mixed with social sciences and other fields (e.g., computing technology for music), so text mining may still pick out hot topics within these UoAs. Whilst topics easily translate into obvious and common keywords, research quality has unknown and probably field dependent translation into research quality (e.g., "improved accuracy" [computing] vs. "surprising connection" [humanities]). Thus, text-based predictions of quality are likely to leverage topic-relevant keywords and perhaps methods as indirect indicators of quality rather than more subtle textual expressions of quality. It is not clear whether input sets that include both citations and text information would leverage hot topics from the text, since the citations would point to the hot topics anyway. Similarly, AI applied to REF articles may identify the topics or methods of the best groups and learn to predict REF scores from them, which would be accurate but undesirable. Article abstracts were pre-processed with a large set of rules to remove publisher copyright messages, structured abstract headings, and other boilerplate texts.

We also included journal names on the basis that journals are key scientific gatekeepers and that a high average citation impact does not necessarily equate to publishing high quality articles. Testing with and without journal names suggested that their inclusion tended to slightly improve accuracy.

- 11-1000. **Title and abstract word unigrams, bigrams, and trigrams** within sentences (i.e., words and phrases of 2 or 3 words). Feature selection was used (chi squared) to identify the best features, always keeping all Input Set 2 features. **Journal names** are also included, for a total of 990 text inputs, selected from the full set as described below.

## 2.3 Machine learning methods

We used machine learning stages that mirror those of a prior study with mostly the same settings. As previously argued, predictions may leverage bibliometric data and text, the latter on the basis that the formula for good research may be identifiable from a text analysis of abstracts. We used 32 machine learning methods, including classification, regression, and ordinal algorithms (Table 2). These include the methods of the prior study (Thelwall, 2022) and the Extreme Gradient Boosting Classifier, which has recently demonstrated good results (Klemiński et al., 2021). Accuracy was calculated after training on 10%, 25% or 50% of the data and evaluated on the remaining articles. These percentages represent a range of realistic options for the REF. Whilst using 90% of the data for training is standard for machine learning, it is not realistic for the REF. Training and testing was repeated 10 times, reporting the average accuracy.

The main differences between the current study and the prior paper (Thelwall, 2022) are as follows.
- Human REF scores instead of journal rankings (split into thirds).
- An additional AI method, Extreme Gradient Boost.

- Hyperparameter tuning with the most promising AI methods in an attempt to improve their accuracy.
- REF Units of Assessment (UoAs) instead of Scopus narrow fields as the main analysis grouping, although we still used Scopus narrow fields for field normalisation of the citation data (MNLCS and NLCS).
- Additional pre-processing rules to catch boilerplate text caught by the rules used for the previous article but found during the analysis of the results for that article.
- Abstract readability, average journal impact, number of institutional affiliations, and first/maximum author productivity/impact inputs.
- In a switch from an experimental to a pragmatic perspective, we used percentages of the available data as the training set sizes for the algorithms rather than fixed numbers of articles.
- Merged years datasets: we combined the first five years (since these had similar results in the prior study) and all years as well as assessing years individually. The purpose of these is to assess whether grouping articles together can give additional efficiency in the sense of predicting article scores with less training data but similar accuracy.
- Merged subjects datasets: we combined all UoAs within each of the four main panel grouping of UoAs to produce four very broad disciplinary groupings. This assesses whether grouping articles together can give additional efficiency in the sense of predicting article scores with less training data but similar accuracy.
- Active learning (Settles, 2011). We used this in addition to standard machine learning. With this strategy, instead of a fixed percentage of the AI inputs having human scores, the algorithm selects the inputs for the humans to score. First, the system randomly selects a small proportion of the articles and the human scores for them are used to build a predictive model. Next, the system selects another set of articles having predicted scores with the lowest probability of being correct for human scoring. This second set is then added to the AI model inputs to rebuild the AI model, repeating the process until a pre-specified level of accuracy is achieved. For the current article, we used batches of 10% to mirror what might be practical for the REF. Thus, a random 10% of the articles were fed into the AI system, then up to 8 further batches of 10% were added, selected to be the articles with the lowest AP prediction probability. Active learning has two advantages: human coders score fewer of the easy cases that the AI system can reliably predict, and scores for the marginal decisions may help to train the system better.
- Correlations are reported.

The set of inputs is not exhaustive because many others have been proposed. We excluded previously used inputs for the following reasons: peer review reports (Li et al., 2019b) because few are public; topic models built from article text (Chen & Zhang, 2015) since this seems unnecessarily complex for large scale application; citation count time series (Abrishami & Aliakbary, 2019) due to not being relevant enough for quality prediction; citation network structure (Zhao & Feng, 2022) since this was not available and is not relevant enough for quality prediction; language (Su, 2020) because most UK articles are English. Other excluded inputs, and corresponding reason for exclusion, were: funding organisation (Su, 2020) because funding is very diverse across the REF and the information was not available; research methods and study details (Jones & Alam, 2019) because full text was not available for most articles; semantic shifts in terms (Tan et al., 2020) because this is overcomplex for the task, although it seems promising; altmetrics (Akella et al., 2021) since these can be manipulated; specific title features, such as title length or the presence of colons (Van Wesel et al., 2014) because these seem too superficial, minor and with varied results.

Table 2. Machine learning methods chosen for regression and classification. Those marked with /o have an ordinal version.

| Code | Method | Type |
| --- | --- | --- |
| bnb/o | Bernoulli Naive Bayes | Classifier |
| cnb/o | Complement Naive Bayes | Classifier |
| gbc/o | Gradient Boosting Classifier | Classifier |
| xgb/o | Extreme Gradient Boosting Classifier | Classifier |
| knn/o | k Nearest Neighbours | Classifier |
| lsvc/o | Linear Support Vector Classification | Classifier |
| log/o | Logistic Regression | Classifier |
| mnb/o | Multinomial Naive Bayes | Classifier |
| pac/o | Passive Aggressive Classifier | Classifier |
| per/o | Perceptron | Classifier |
| rfc/o | Random Forest Classifier | Classifier |
| rid/o | Ridge classifier | Classifier |
| sgd/o | Stochastic Gradient Descent | Classifier |
| elnr | Elastic-net regression | Regression |
| krr | Kernel Ridge Regression | Regression |
| lasr | Lasso Regression | Regression |
| lr | Linear Regression | Regression |
| ridr | Ridge Regression | Regression |
| sgdr | Stochastic Gradient Descent Regressor | Regression |

# 3 Results

## 3.1 RQ1, RQ2: Primary AI prediction accuracy tests

The accuracy of each machine learning method was calculated for each year range (2014, 2015, 2016, 2017, 2018, 2019, 2020, 2014-18, 2014-20), separately by UoA and Main panel. The results are reported as accuracy above the baseline (accuracy-baseline)/(1-baseline), where the baseline is the proportion of articles with the most common score (usually 4* or 3*). The results are reported only for 2014-18 combined, with the graphs for the other years available online, as are graphs with 10% or 25% training data (http://cybermetrics.wlv.ac.uk/TechnologyAssistedResearchAssessment.html). The overall level of accuracy for each individual year from 2014 to 2018 tended to be similar, with lower accuracy for 2019 and 2020 due to the weaker citation data. Combining 2014 to 2018 gave a similar level of accuracy to that of the individual years, and so it is informative to focus on this set. With the main exception of UoA 8 Chemistry, the accuracy of the machine learning methods was higher with 1000 inputs (input set 3) than with 9 or 10 (input sets 1 or 2), so only the results for the largest set are reported.

Articles 2014-18 in most UoAs could be classified with above baseline accuracy with at least one of the tested AI methods, but there are substantial variations between UoAs (Figure 1). There is not a simple pattern in terms of the types of UoA that are easiest to classify. This is partly due to differences in sample sizes and probably also affected the variety of the fields within each UoA (e.g., Engineering is a relatively broad UoA compared to Archaeology). Seven UoAs had accuracy at least 0.3 above the baseline, and these are from the health and physical sciences as well as UoA 16: Economics and Econometrics. Despite this variety, the level of AI accuracy is very low for all Main Panel D (mainly arts and humanities) and for most of Main Panel C (mainly social sciences).

Although larger sample sizes help the training phase of machine learning, the UoA with the most articles (12: Engineering) had only moderate accuracy, so the differences between UoAs are also partly due to differing underlying AI prediction difficulties between fields.

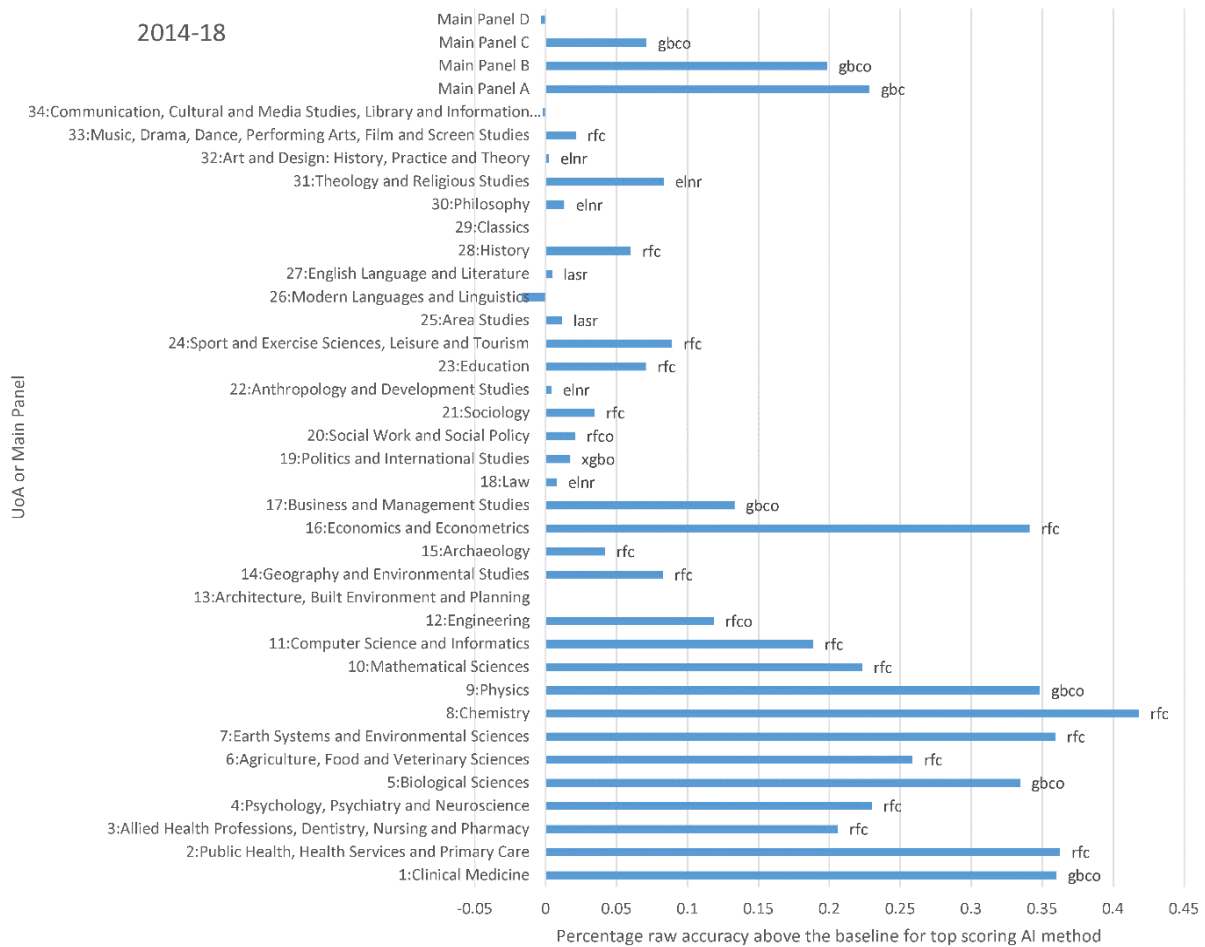

Figure 1. The percentage accuracy above the baseline for the most accurate machine learning method, trained on **50%** of the 2014-18 Input Set 3: Bibliometrics, journal impact and text, after excluding articles with shorter than 500-character abstracts **and excluding duplicate articles within each UoA**. No models were built for Classics due to too few articles. Average across 10 iterations.

The most accurate UoAs are not all the same as those with highest accuracy above the baseline because there were substantial differences in the baselines between UoAs (Figure 2). The predictions were up to 72% accurate (UoAs 8,16), with 12 UoAs having accuracy above 60%. The lowest raw accuracy was 46% (UoA 23). If accuracy is assessed in terms of correlations, then the AI predictions always positively correlate with the human scores, but at rates varying between 0.0 (negligible) to 0.6 (strong) (Table 3). These correlations roughly match the prediction accuracies. Note that the correlations are much higher when aggregated by institution, reaching 0.998 for total institutional scores in UoA 1 (Table 4).

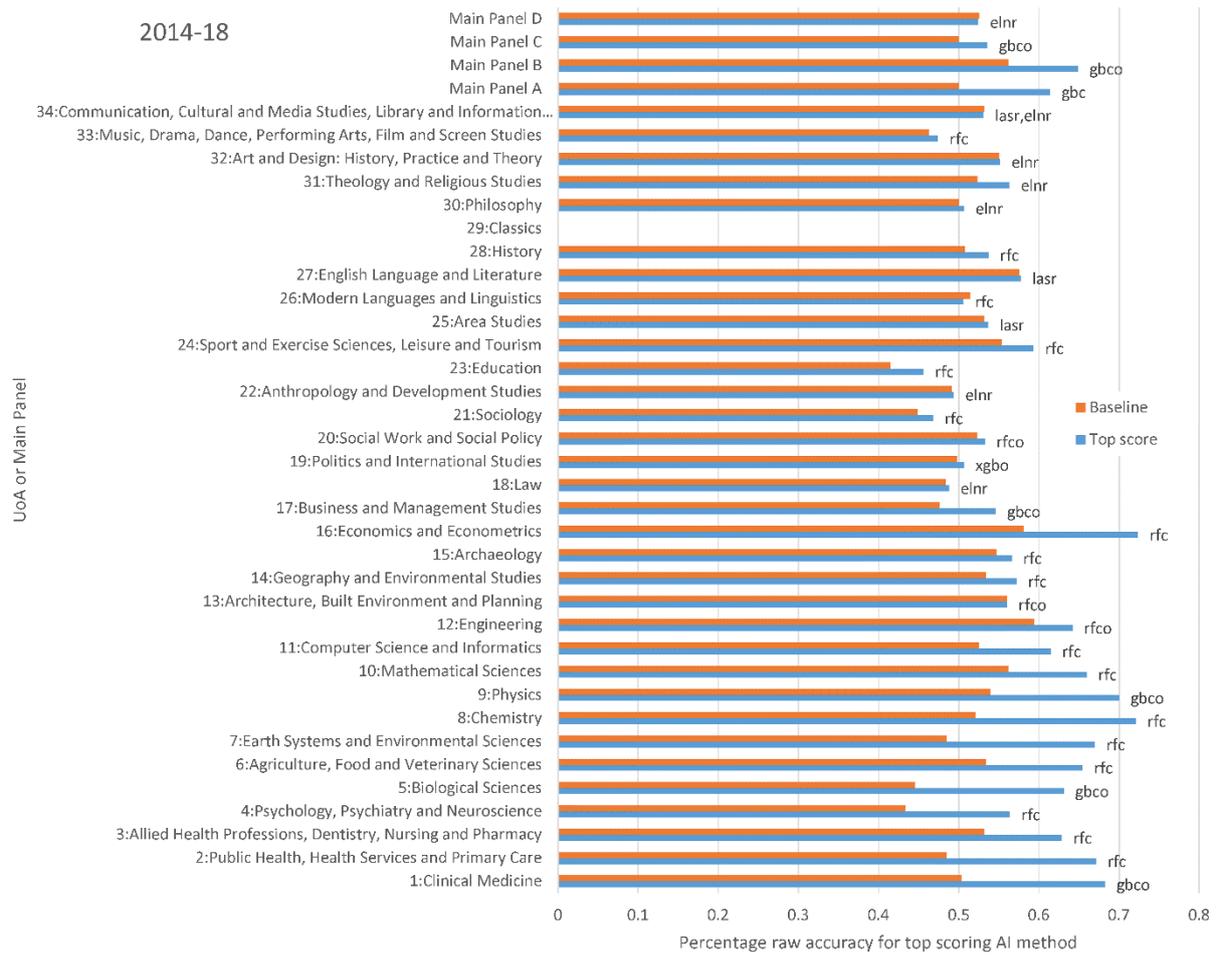

Figure 2. As for the previous figure but showing raw accuracy.

Table 3. Pearson correlations between AI predictions and actual scores for articles 2014-18 with 50% used for training, following Strategy 1 (averaged across 10 iterations).

| Dataset | Articles 2014-18 | Predicted at 50% | Correlation |
|---|---|---|---|
| 1:Clinical Medicine | 7274 | 3637 | 0.562 |
| 2:Public Health, Health Services and Primary Care | 2855 | 1427 | 0.507 |
| 3:Allied Health Professions, Dentistry, Nursing and Pharmacy | 6962 | 3481 | 0.406 |
| 4:Psychology, Psychiatry and Neuroscience | 5845 | 2922 | 0.474 |
| 5:Biological Sciences | 4728 | 2364 | 0.507 |
| 6:Agriculture, Food and Veterinary Sciences | 2212 | 1106 | 0.452 |
| 7:Earth Systems and Environmental Sciences | 2768 | 1384 | 0.491 |
| 8:Chemistry | 2314 | 1157 | 0.505 |
| 9:Physics | 3617 | 1808 | 0.472 |
| 10:Mathematical Sciences | 3159 | 1579 | 0.328 |
| 11:Computer Science and Informatics | 3292 | 1646 | 0.382 |
| 12:Engineering | 12511 | 6255 | 0.271 |
| 13:Architecture, Built Environment and Planning | 1697 | 848 | 0.125 |
| 14:Geography and Environmental Studies | 2316 | 1158 | 0.277 |
| 15:Archaeology | 371 | 185 | 0.283 |
| 16:Economics and Econometrics | 1083 | 541 | 0.511 |
| 17:Business and Management Studies | 7535 | 3767 | 0.353 |
| 18:Law | 1166 | 583 | 0.101 |
| 19:Politics and International Studies | 1595 | 797 | 0.181 |
| 20:Social Work and Social Policy | 2045 | 1022 | 0.259 |
| 21:Sociology | 949 | 474 | 0.180 |
| 22:Anthropology and Development Studies | 618 | 309 | 0.040 |
| 23:Education | 2081 | 1040 | 0.261 |
| 24:Sport and Exercise Sciences, Leisure and Tourism | 1846 | 923 | 0.265 |
| 25:Area Studies | 303 | 151 | 0.142 |
| 26:Modern Languages and Linguistics | 630 | 315 | 0.066 |
| 27:English Language and Literature | 424 | 212 | 0.064 |
| 28:History | 583 | 291 | 0.141 |
| 29:Classics | 56 | 0 | - |
| 30:Philosophy | 426 | 213 | 0.070 |
| 31:Theology and Religious Studies | 107 | 53 | 0.074 |
| 32:Art and Design: History, Practice and Theory | 665 | 332 | 0.028 |
| 33:Music, Drama, Dance, Performing Arts, Film and Screen Studies | 350 | 175 | 0.164 |
| 34:Communication, Cultural and Media Studies, Library and Information Management | 583 | 291 | 0.084 |

Table 4. Pearson correlations between AI predictions and actual scores for articles 2014-18 with 50% used for training, following Strategy 1 (averaged across 10 iterations) and aggregated by institution for UoAs 1-11 and 16.

| UoA | Actual vs AI predicted average score | Actual vs AI predicted total score |
|---|---|---|
| 1:Clinical Medicine | 0.895 | 0.998 |
| 2:Public Health, Health Services and Primary Care | 0.906 | 0.995 |
| 3:Allied Health Professions, Dentistry, Nursing & Pharmacy | 0.747 | 0.982 |
| 4:Psychology, Psychiatry and Neuroscience | 0.844 | 0.995 |
| 5:Biological Sciences | 0.885 | 0.995 |
| 6:Agriculture, Food and Veterinary Sciences | 0.759 | 0.975 |
| 7:Earth Systems and Environmental Sciences | 0.840 | 0.986 |
| 8:Chemistry | 0.897 | 0.978 |
| 9:Physics | 0.855 | 0.989 |
| 10:Mathematical Sciences | 0.664 | 0.984 |
| 11:Computer Science and Informatics | 0.724 | 0.945 |
| 16:Economics and Econometrics | 0.862 | 0.974 |

Hyperparameter tuning systematically searches a range of input parameters for machine learning algorithms, looking for variations that improve their accuracy. Whilst this marginally increases accuracy on some UoAs it marginally reduces it on others, so has little difference overall (Figure 3).

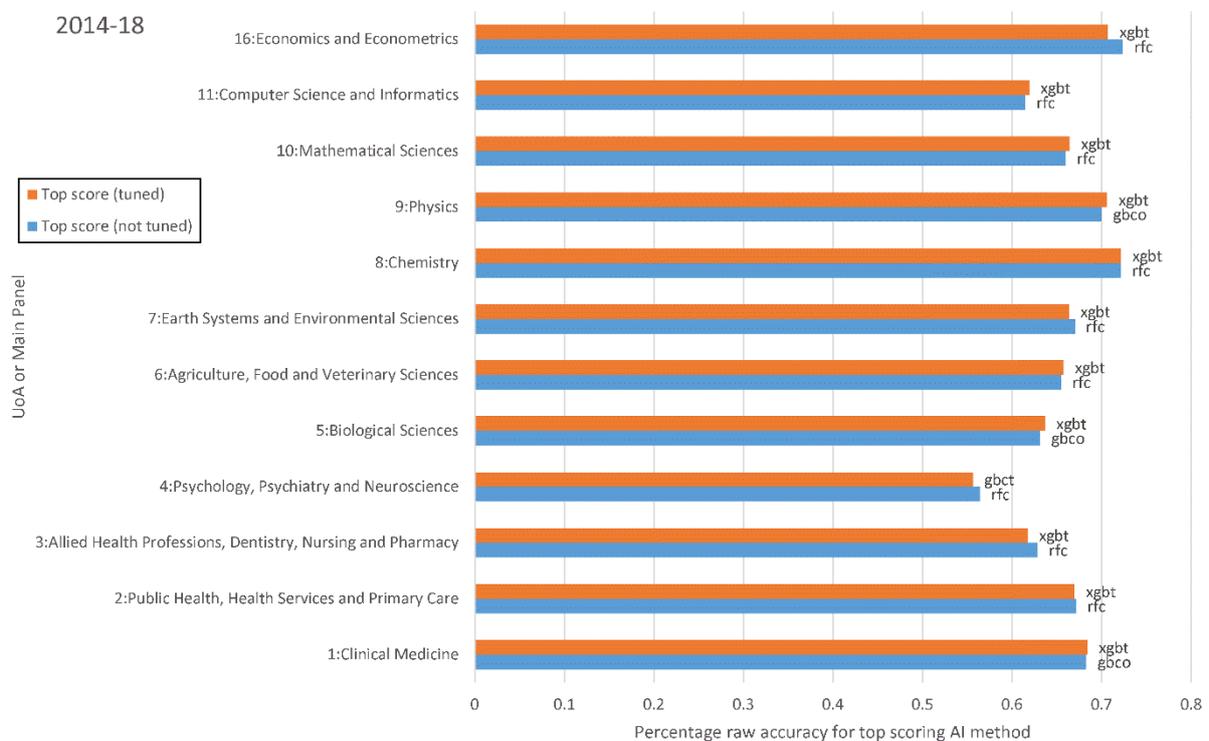

Figure 3. The percentage accuracy for the most accurate machine learning method with and without hyperparameter tuning (out of the main six), trained on **50%** of the 2014-18 articles and **Input set 3: bibliometrics + journal impact + text; 1000 features in total**. The most accurate method is named.

## 3.2 RQ3 High prediction probability subsets

The methods used to predict article scores report an estimate of the probability that these predictions are correct. If these estimates are not too inaccurate, then arranging the articles in descending order prediction probability can be used to identify subsets of the articles that can have their REF score estimated more accurately than for the set overall.

The graphs below for the UoAs with the most accurate predictions (Figure 4) can be used to read the number of scores that can be predicted with any given degree of accuracy. For example, setting the accuracy threshold at 90%, 500 articles could be predicted in UoA 1. The graphs report the accuracy by comparison with sub-panel provisional scores rather than the AI accuracy estimates. The graphs confirm that higher levels of AI score prediction accuracy can be obtained for subsets of the predicted articles. Nevertheless, they suggest that there is a limit to which this is possible. For example, no UoA can have substantial numbers of articles predicted with accuracy above 95% and UoA 11 has few articles that can be predicted with accuracy above 80%.

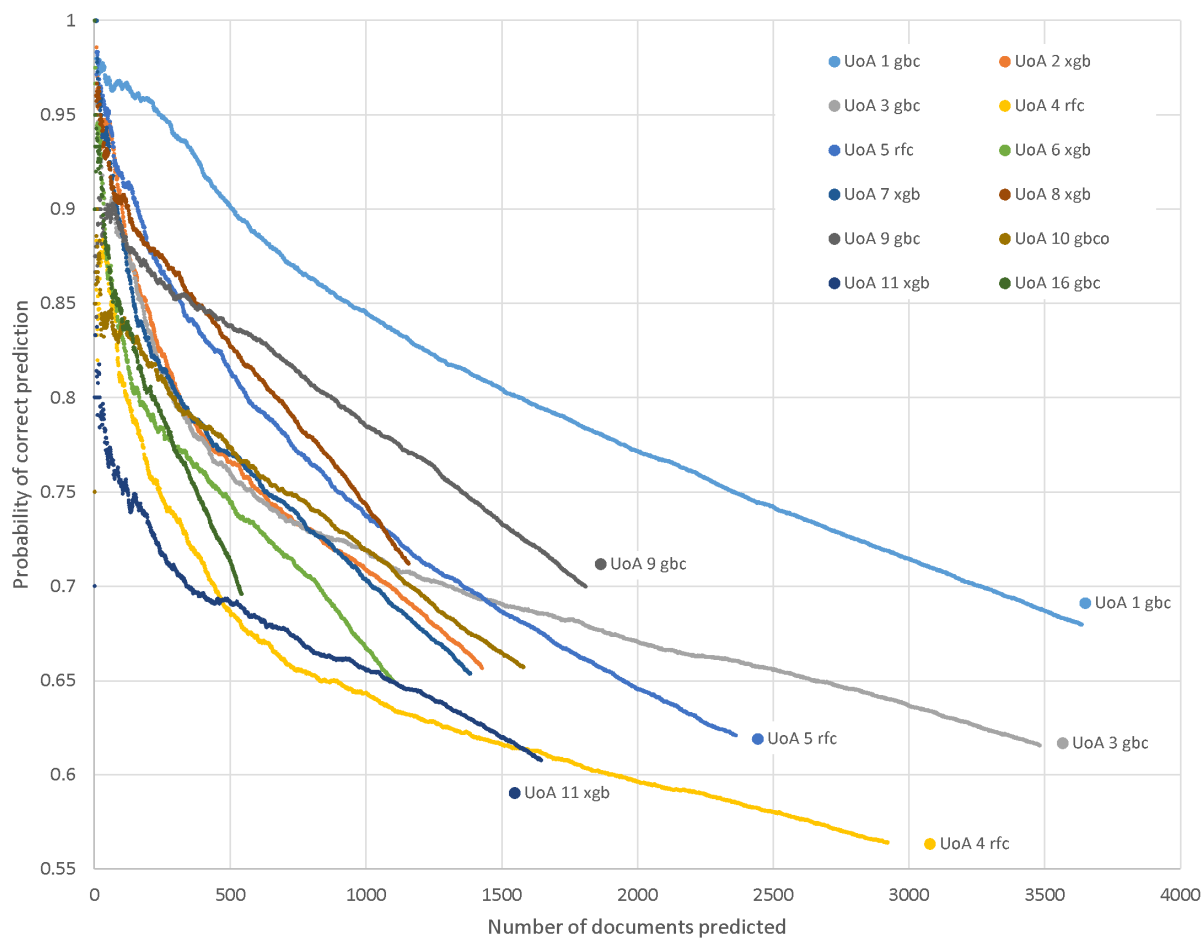

Figure 4. Probability of an AI prediction (best machine learning method at the 85% level, trained on 50% of the data 2014-18 with 1000 features) being correct against the number of predictions for UoAs 1-11, 16. The articles are arranged in order of the probability of the prediction being correct, as estimated by the AI. Each point is the average across 10 separate experiments.

If the algorithm is trained on a lower percentage of the articles, then fewer scores can be predicted at any high level of accuracy, as expected (Figure 5).

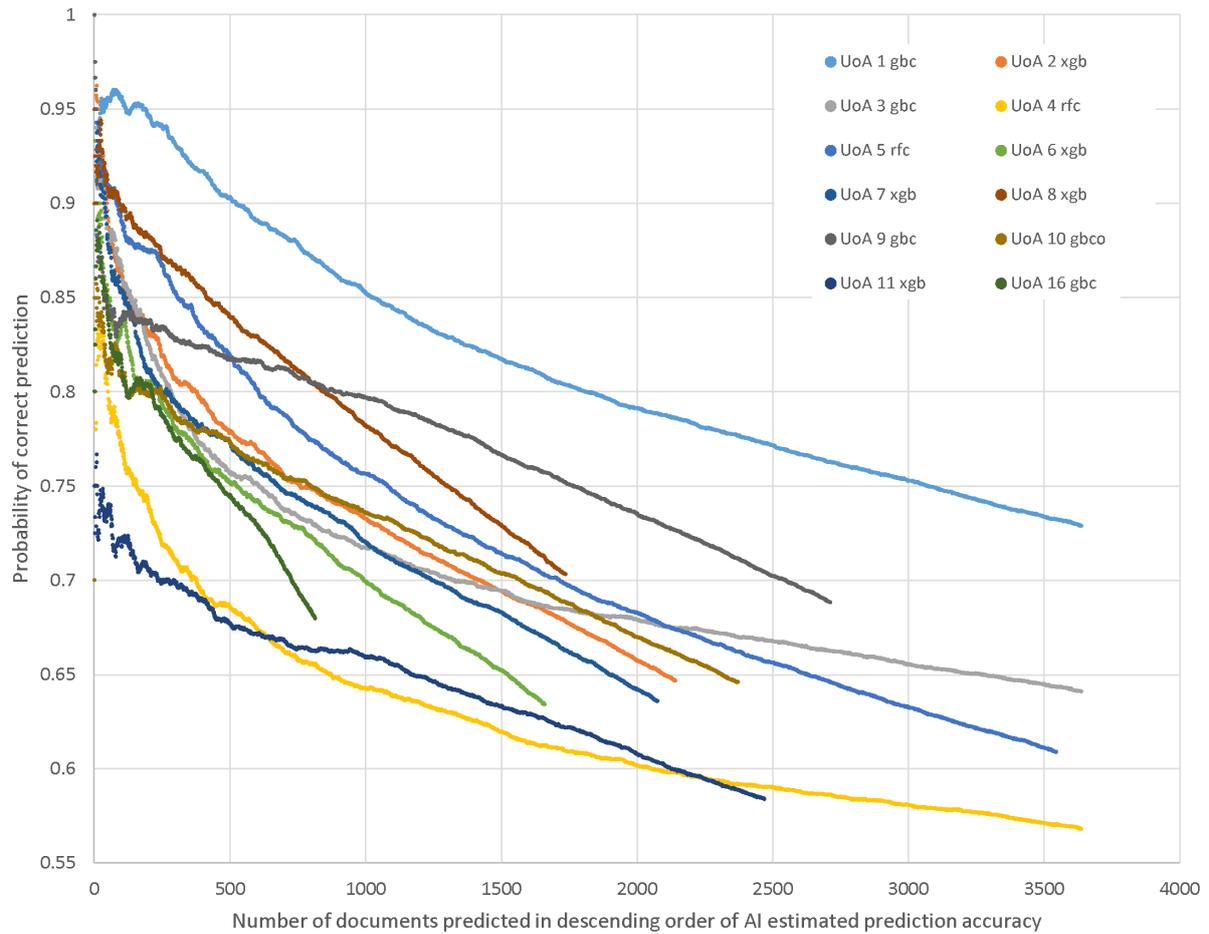

Figure 5. As above but trained on 25% of the data.

## 3.3 Active learning summary

The active learning strategy, like that of selecting high prediction probability scores, is successful at generating higher prediction probability subsets (Figure 6). Active learning works better for some UoAs relative to others, and is particularly effective for UoAs 1, 4, and 5 in the sense that their accuracy increases faster than the others as the number training set size increases.

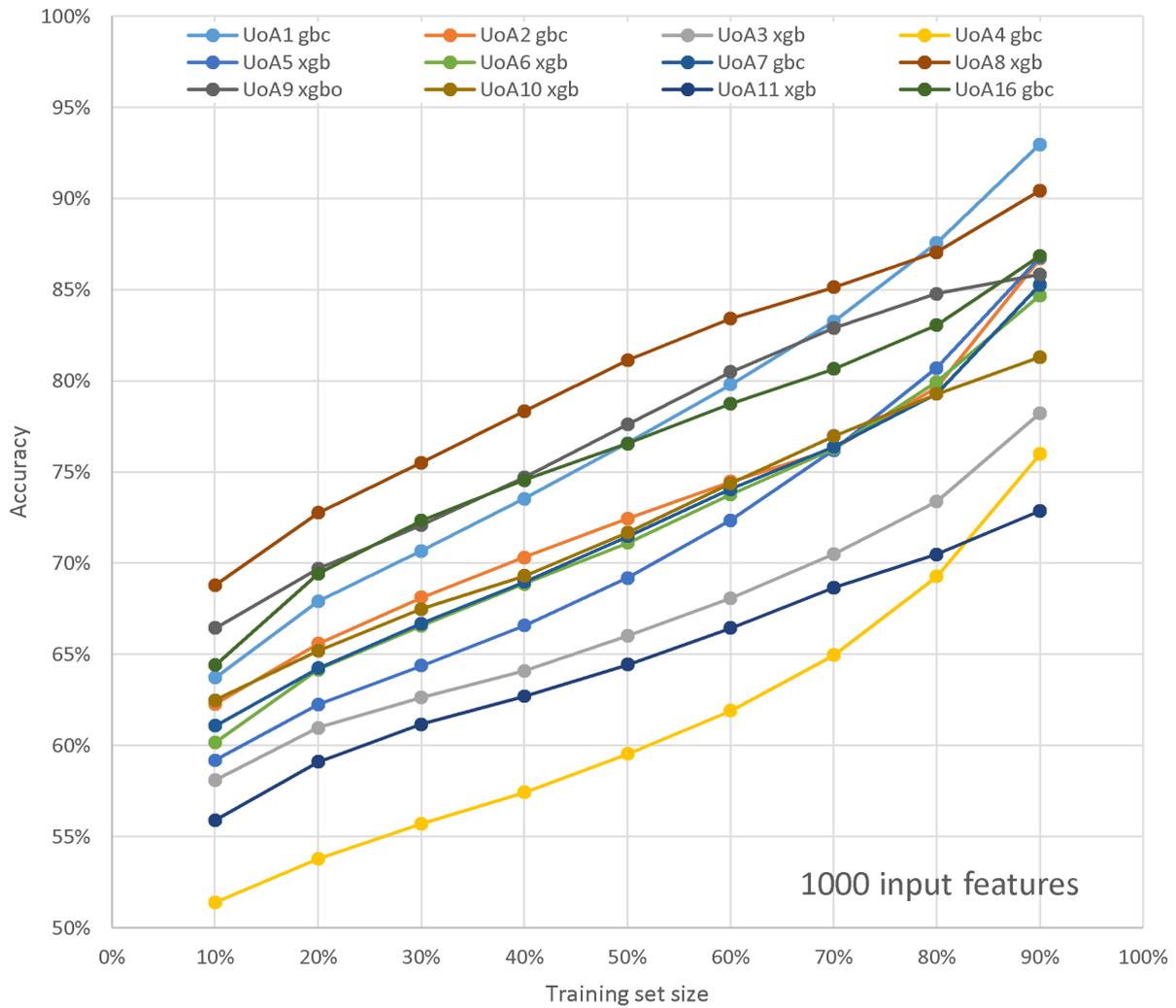

Figure 6. Active learning on UoAs 1-11, 16 showing the results for the machine learning method with the highest accuracy at 90% and 1000 input features. Results are the average of 40 independent full active learning trials.

Active learning overall can predict more articles at realistic thresholds than the high prediction probability strategy (Table 5). Here, 85% is judged to be a realistic accuracy threshold as a rough estimate of human-level accuracy. In the twelve highest prediction probability UoAs, active learning identifies more articles (3688) than the high prediction probability strategy (2879) and a higher number in all UoAs where the 85% threshold is reached. Active learning is only less effective when the threshold is not reached.

Table 5. The Number of articles that can be predicted at an accuracy above **85%** using active learning or high prediction probability subsets in UoAs 1-11,16. Overall accuracy includes the human scored texts for eligible and ineligible articles.

| UoA | Human scored articles | Human scored articles % | Active learning accuracy | AI predicted articles | High prediction probability articles |
|---|---|---|---|---|---|
| 1:Clinical Medicine | 5816 | 80% | 87.6% | 1458 | 952* |
| 2:Public Health, Health Serv. & Primary Care | 2565 | 90% | 86.7% | 290 | 181 |
| 3:Allied Health Prof., Dentist., Nurs. Pharm. | 6962 | 100% |  | 0 | 163 |
| 4:Psychology, Psychiatry & Neuroscience | 5845 | 100% |  | 0 | 66 |
| 5:Biological Sciences | 4248 | 90% | 86.8% | 480 | 308 |
| 6:Agriculture, Food & Veterinary Sciences | 2212 | 100% |  | 0 | 86 |
| 7:Earth Systems & Environmental Sciences | 2484 | 90% | 85.3% | 284 | 142 |
| 8:Chemistry | 1617 | 70% | 85.1% | 697 | 402* |
| 9:Physics | 3249 | 90% | 85.9% | 368 | 362 |
| 10:Mathematical Sciences | 3159 | 100% |  | 0 | 86 |
| 11:Computer Science & Informatics | 3292 | 100% |  | 0 | 29 |
| 16:Economics & Econometrics | 972 | 90% | 86.9% | 111 | 102 |
| Total |  |  |  | 3688 | 2879 |

* 25% training set size instead of 50% training set size because more articles were predicted.

### 3.4 RQ4: HEI-level accuracy

For the UK REF, as for other national evaluation exercises, the most important unit of analysis is the institution because the results are used to allocate funding (or a pass/fail decision) to institutions for a subject rather than to individual articles or researchers. At the institutional level, there can be non-trivial score shifts for individual institutions, even with high prediction probabilities. UoA 1 has one of the lowest average score shifts due to relatively large institutional sizes, but these are still non-trivial (Figure 7). The score shifts are largest for small institutions, for statistical reasons, but there is also a degree of bias, in the sense of institutions that then to benefit or lose out overall from the AI predictions. The biggest score shift for a relatively large number of articles in a UoA (one of the five largest sets in the UoA) is 11% (UoA 7) or 1.9% overall (UoA 8), considering 100% accuracy for the articles given human scores (Table 6). Whilst 1.9% is a small percentage, it may represent the salaries of multiple members of staff and so is a non-trivial consideration. The institutional score shifts are larger for Strategy 1 (not shown).

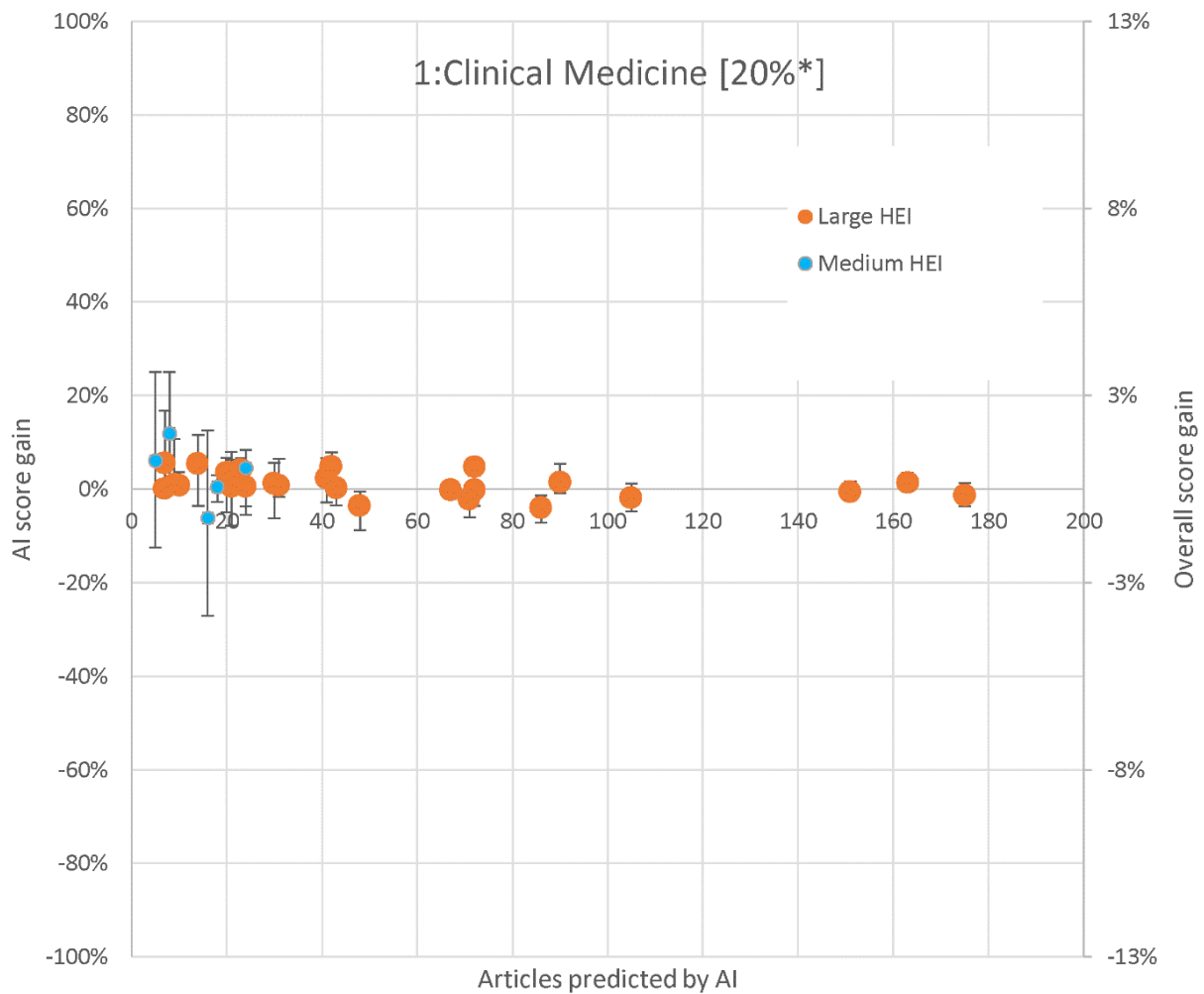

Figure 7. The average REF AI institutional score gain on UoA 1: Clinical Medicine for the most accurate machine learning method with active learning, stopping at 85% accuracy on the 2014-18 data and **bibliometric + journal + text inputs, after excluding articles with shorter than 500 character abstracts**. AI score gain is a financial calculation (4*=100% funding, 3*=25% funding, 0-2*=0% funding). The x axis records the number of articles with predicted scores in one of the iterations. The right-hand axis shows the overall score gain for all REF journal articles, included those that would not be predicted by AI. Error bars indicate the highest and lowest values from 10 iterations.

Table 6. Maximum average AI score shifts for five largest Higher Educational Institution (HEI) submissions and for all HEI submissions with active learning with an 85% threshold. The same information for the largest AI score shifts rather than the average score shifts. Overall figures include all human coded journal articles.

| UoA or Panel | Human scores % | Max HEI av. score shift (overall) | Max top 5 HEIs av. score shift (overall) | Max HEI largest score shift (overall) | Max top 5 HEIs largest score shift (overall) |
|---|---|---|---|---|---|
| 1:Clinical Medicine | 80% | 12% (1.5%) | 1.9% (0.2%) | 27% (3.4%) | 5.4% (0.7%) |
| 2:Public Health, H. Services & Primary Care | 90% | 27% (1.7%) | 13% (0.8%) | 75% (4.7%) | 16% (1.0%) |
| 3:Allied Health Prof., Dentist Nurs Pharm | 100% | | | | |
| 4:Psychology, Psychiatry & Neuroscience | 100% | | | | |
| 5:Biological Sciences | 90% | 63% (3.9%) | 7.3% (0.5%) | 75% (4.7%) | 10% (0.6%) |
| 6:Agriculture, Food & Veterinary Sciences | 100% | | | | |
| 7:Earth Systems & Environmental Sciences | 90% | 32% (2.0%) | 11% (0.7%) | 75% (4.7%) | 16% (1.0%) |
| 8:Chemistry | 70% | 11% (2.1%) | 10% (1.9%) | 75% (14%) | 14% (2.6%) |
| 9:Physics | 90% | 10% (0.6%) | 3.7% (0.2%) | 75% (4.7%) | 10% (0.6%) |
| 10:Mathematical Sciences | 100% | | | | |
| 11:Computer Science & Informatics | 100% | | | | |
| 16:Economics and Econometrics | 90% | 35% (2.2%) | 5.1% (0.3%%) | 75% (4.7%) | 19% (1.2%) |

Bias occurs in the predictions from active learning, even at a high level of accuracy. For example, in most UoAs, larger HEIs, HEIs with higher average scores, and HEIs submitting more articles to a UoA tend to be disadvantaged by AI score predictions (Figure 8). This is not surprising because, other factors being equal, high scoring HEIs would be more likely to lose from an incorrect score prediction. This is because they would have a higher proportion of top scoring articles (which would always be downgraded by errors). Similarly, larger HEIs tend to submit more articles and tend to have higher scores.

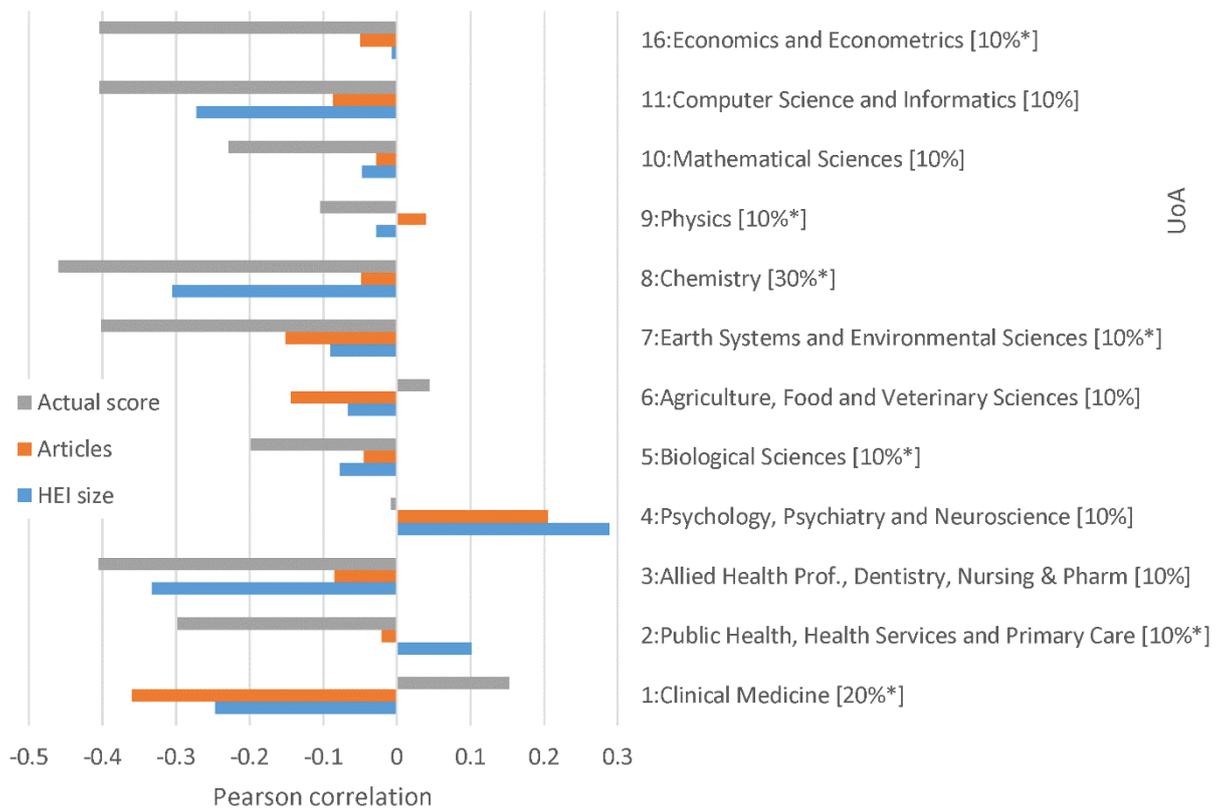

Figure 8. Pearson correlations between institutional size (number of articles submitted to REF) or submission size (number of articles submitted to UoA) or average institutional REF score for the UoA and average REF AI institutional score gain on UoA 1:Clinical Medicine to UoA 16: Economics and Econometric for the most accurate machine learning method with active learning, stopping at 85% accuracy on the 2014-18 data and **bibliometric + journal + text inputs, after excluding articles with shorter than 500 character abstracts**. Captions indicate the proportion of journal articles predicted, starred if the 85% accuracy active learning threshold is met.

## 3.5 RQ5: Accuracy on Scopus broad fields

If the REF articles are organised into Scopus broad fields before classification, then the most accurate machine learning method is always gbco, rfc, rfco or xgbo. The highest accuracy above the baseline is generally much lower in this case than for the REF fields, with only Multidisciplinary having accuracy above the baseline above 0.3, with the remainder being substantially lower (Figure 9). The lower accuracy is because the Scopus broad fields are effectively much broader than UoAs. They are journal-based rather than article-based and journals can be allocated multiple categories. Thus, a journal containing medical engineering articles might be found in both the Engineering and the Medicine categories. This interdisciplinary, broader nature of Scopus broad fields reduces the accuracy of the machine learning methods, despite the field normalised indicators used in them.

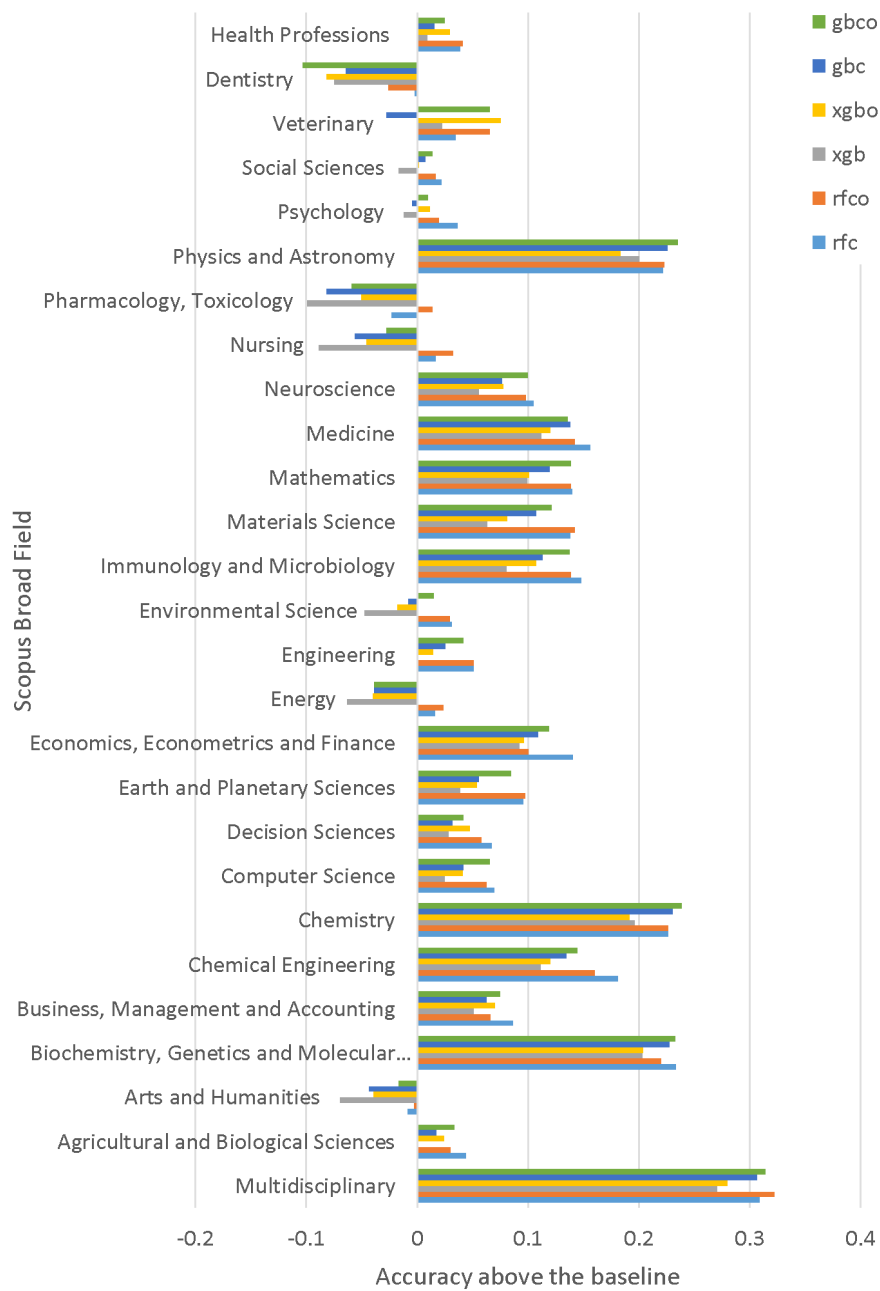

Figure 9. The percentage accuracy above the baseline on Scopus broad fields for the three most accurate machine learning methods and their ordinal variants, trained on **50%** of the 2014-18 Input Set 3: Bibliometrics, journal impact and text, after excluding articles with shorter than 500-character abstracts, zero scores or duplicate within a Scopus broad field.

## 4 Discussion

The results are limited to articles from a single country and period. These articles are self-selected as presumably the best works (1 to 5 per person) of the submitting UK academics over the period 2014-2020. The findings used three groups (1*-2*, 3*,4*) and finer grained outputs (e.g., the 27-point Italian system, 3-30) would be much harder to predict. The results are also limited by the scope of the UoAs examined. AI predictions for countries with less Scopus-indexed work to analyse, or with more recent work, would probably be less accurate. The results may also change in the future as the scholarly landscape evolves, including journal formats, article formats, and citation practices. The accuracy statistics may be slightly optimistic due to overfitting: running multiple tests and reporting the best results. This has been mitigated by generally selecting strategies that work well for most UoAs, rather than customising strategies for UoAs. The main source of overfitting is probably the main AI algorithm selection, since six similar algorithms tended to perform well and only the most accurate one for each UoA is reported.

The results generally confirm previous studies in that the inputs used can generate above-baseline accuracy predictions, and that there are substantial disciplinary differences in the extent to which article quality (or impact) can be predicted. The accuracy levels achieved here are much lower than previously reported for attempts to identify high impact articles, however. The accuracy levels are also lower than for the most similar prior study, which predicted journal thirds as a simple proxy for article quality, despite using less training data in some cases and a weaker set of inputs (Thelwall, 2022). Collectively, this suggests that the task of predicting article quality is substantially harder than the task of predicting article citation impact. Presumably this is due to high citation specialties not always being high quality specialties.

Some previous studies have used input features extracted from article full texts for collections of articles where this is easily available. To check whether this is a possibility here, the full text of 59,194 REF-submitted articles was supplied from the core.ac.uk repository of open access papers (Knoth & Zdrahal, 2012) by Prof Petr Knoth, Maria Tarasiuk and Matteo Cancellieri. These matched 43.3% of the REF articles with scores and strategy 1 was rerun with this reduced set, incorporating word counts, character counts, figure counts, and table counts extracted from the full text, but accuracy was lower. This was probably partly due to the full texts often containing copyright statements and line numbers as well as occasional scanning errors, and partly due to the smaller training set sizes. Tests of other suggested features (supplementary materials, data access statements) found very low correlations with scores, so these were not included.

# 5 Conclusion

The results show that AI predictions of article quality scores on a three-level scale are possible from article metadata, citation information, author career information and title/abstract text with up to 72% accuracy in some UoAs for articles older than two years, given enough articles for training. Substantially higher levels of accuracy may not be possible due to the need for tacit knowledge to understand the context of articles to properly evaluate their contributions. Whilst academic impact can be directly assessed to some extent through citations, robustness and originality are difficult to assess from citations and metadata, although journals may be partial indicators of these in some fields and original themes can in theory be detected (Chen, et al., 2022). The tacit knowledge needed to assess the three components of quality may be more important in fields (UoAs) in which lower AI prediction accuracy was attained. Higher accuracy may be possible with other inputs included, such as article full text (if cleaned and universally available) and peer reviews (if widely available).

The results suggest, somewhat surprisingly, that random forest classifier and the gradient boosting classifier tend to be the most accurate (both classification and ordinal variants) rather than the extreme gradient boosting classifier. Nevertheless, xgb is the most accurate for some UoAs, and especially if active learning is used.

If high levels of accuracy are needed, small subsets of articles can be identified in some UoAs that can be predicted with accuracy above a given threshold through active learning.

This could be used when phased peer review is practical, so that initial review scores could be used to build predictions and a second round of peer review would classify articles with low probability AI predictions. Even at relatively high prediction probability levels, AI predictions can shift the scores of small institutions substantially due to statistical variations and larger institutions due to systematic biases in the AI predictions, such as against high scoring institutions.

Although it would be possible to use the models built for the current paper in the next REF (probably 2027), their accuracy could not be guaranteed. For example, the text component would not reflect newer research topics (e.g., any future virus) and future citation data would not be directly comparable since average citation counts may continue to increase in future years.

Finally, unless new AI approaches can be found, it seems clear that AI prediction of scores is irrelevant for the arts and humanities, most of the social sciences and engineering, and is weak for some of the remaining areas.

**AUTHOR CONTRIBUTIONS**
Mike Thelwall: Methodology, Writing–original draft, Writing–review & editing.
Kayvan Kousha: Methodology, Writing–review & editing.
Paul Wilson: Methodology, Writing–review & editing.
Meiko Makita: Methodology, Writing–review & editing.
Mahshid Abdoli: Methodology, Writing–review & editing.
Emma Stuart: Methodology, Writing–review & editing.
Jonathan Levitt: Methodology, Writing–review & editing.
Petr Knoth: Methodology, Data curation.
Maria Tarasiuk: Data curation.
Matteo Cancellieri: Data curation.
**COMPETING INTERESTS**
The authors have no competing interests to declare.
**FUNDING INFORMATION**
This study was funded by Research England, Scottish Funding Council, Higher Education Funding Council for Wales, and Department for the Economy, Northern Ireland as part of the Future Research Assessment Programme (https://www.jisc.ac.uk/future-research-assessment-programme). The content is solely the responsibility of the authors and does not necessarily represent the official views of the funders.
**DATA AVAILABILITY**
Extended versions of the results are available in the full report (http://cybermetrics.wlv.ac.uk/TechnologyAssistedResearchAssessment.html). The raw data was deleted before submission to follow UKRI policy for REF2021.